\documentstyle[12pt]{article}
\newcommand{\btau}{\mbox{\boldmath $\tau $}}
\newcommand{\bpi} {\mbox{\boldmath $\pi  $}}

\begin{document}
\title{CHIRAL PHASE TRANSITIONS}
\author{
{M. D. Scadron} 
\begin{footnote}
{Permanent address: Physics Dept., University of Arizona, Tucson, AZ, USA}
\end{footnote}
\begin{footnote}
{e-mail: scadron@physics.arizona.edu}
\end{footnote}
and {P. \.Zenczykowski}
\begin{footnote}{e-mail: zenczyko@iblis.ifj.edu.pl}
\end{footnote}
\\
\\
{\em Department of Theoretical Physics,} \\
{\em Institute of Nuclear Physics,}\\
{\em Krak\'ow, Poland}\\
}
\maketitle
\begin{abstract}
We show that melting the quark mass, the scalar $\sigma $ mass and the quark
condensate leads uniquely to the quark-level SU(2) linear $\sigma $ model field
theory. Upon thermalization, the chiral phase transition curve requires
$T_c =2f^{CL}_{\pi }\approx 180~ MeV$ when $\mu = 0$, while the critical
chemical potential is $\mu _c =m_q\approx 325~MeV$. 
Transition to the superconductive
phase occurs at $T^{(SC)}_c=\Delta /\pi e^{-\gamma _E}$.
Coloured diquarks suggest $T_c^{(SC)}<180~MeV$.
\end{abstract}
\vfill
Invited talk at the Chiral Fluctuations in Hadronic Matter International
Workshop, Orsay, France, Sept. 2001
\newpage
\section{Introduction}
\label{Intr}

According to the now prevailing views,
it is expected that at high temperature and/or density
the standard hadronic matter
will undergo a phase transition to a new state.
Such prospects have motivated present experimental searches for
the signatures of the quark-gluon plasma in high-energy 
heavy ion collisions.
The appearance of such state(s) 
may also affect properties of strongly
interacting matter at such extreme conditions as those occurring
inside of neutron stars.
Clearly, if we are to understand the underlying phenomena better,
various theoretical studies should be pursued.

Temperature dependence of the properties of low density QCD has been 
 explored fairly well. The basis for our understanding is provided by
 lattice field theory calculations which, for QCD with two massive
 flavours, exhibit a phase transition to a chirally symmetric phase
 at $T_c = 173 \pm 8 ~MeV$ for im\-proved staggered fermions
 \cite{lattice}.
 Dependence on baryon charge density or chem\-ical potential $\mu$ is
 much more hypothetical, since lattice Monte Carlo techniques
 are not well suited to the case when $\mu $ is non-zero.
 Nonetheless, on the basis of general arguments
 and of an insight gained from various models,
 one may conjecture 
 the structure of the phase diagram in QCD with two massless
 quarks.
 
Thus, the salient features of this QCD phase diagram in 
the $T\mu$ plane
are known or expected as follows (see Fig.1 \cite{phasediagram}).\\
\noindent
i) When temperature $T$ is raised at $\mu =0$, 
the chiral QCD condensate
$\langle \bar{q}q \rangle $ melts at $T_c \approx 173\pm 8~MeV$ 
or so, and the
system enters into a quark-gluon plasma (QGP)
phase, where the chiral condensate is identically zero.
The transition from the low temperature hadronic phase to the high
temperature QGP phase is most probably of second order. Because
the QGP and hadronic phases differ by the expectation value of 
$\langle \bar{q}q \rangle $, 
a (curved) line of phase transitions must begin at $T=T_c$,
$\mu =0$ and continue for $\mu > 0$ without terminating. 
By continuity  for low $\mu $,
 these phase transitions must be of second order 
 (this is indicated by a dotted line in Fig. 1).\\
\noindent
ii) When the chemical potential $\mu $ of the system is increased at zero
temperature, one first encounters a transition from the phase of zero
number density to that of nuclear matter density. This transition is of
first order and by continuity remains such also at low $T$.
In Fig. 1 this is indicated by a solid line. 
 The hadronic gas at low $T$ and small $\mu $ has number density
$n(T,\mu) \ne 0$, and is connected analytically to the
nuclear matter phase at high enough $T_0$. The latter is
 related to the binding energy of nucleons in nuclear matter 
 ($T_0 \approx 10~MeV$). \\
\noindent
iii) When the chemical potential and number density are further increased 
at zero temperature, the
interactions of quarks become weak due to (QCD) asymptotic freedom and
colour interactions become screened at shorter and shorter lengths.
Thus, nonperturbative phenomena such as chiral symmetry breaking
should be absent at sufficiently high $\mu $ and a phase
transition should occur at some $\mu _c$. Most likely, it is
of first order. 
Since regions to the left and right of the point $\mu=\mu _c$, $T=0$ 
differ in the
value of the quark condensate, a line
 of first order transitions (marked solid) must start
at this point and continue away from $T=0$ without
terminating. 
Most likely, it joins the line of second order
phase transitions emanating from $\mu =0$, $T=T_c$ at a tricritical
point $T_3$. \\
iv) There are arguments that in the region of low $T$ to the right of
$\mu _c$, one should expect a nonvanishing value of a diquark 
 condensate $\langle qq \rangle $ which spontaneously breaks
 colour symmetry. In this region, for temperatures below 
 $T_c^{(SC)} $ (of order 100 MeV), one expects to find colour
 superconductivity.  The nature of this phase transition to
 a non-superconducting phase at higher temperatures
 is not clear yet and,
 consequently, this transition line is marked with dashes.
 
 In the following it will be shown how some of the above features
  as well as other aspects
  of the underlying theory are predicted in a linear
  sigma model with a built-in dynamical symmetry breaking structure.

\section{Chiral phase transition}
\label{Chphtr}
The word "chiral" means handedness in Greek. Defining left- and right- handed
charges $2 Q_{\pm}^i=Q^i \pm Q^i_5 $, the SU(2) current 
(charge) algebra of the
1960s reduces to $[Q^i_{\pm},Q^j_{\pm}]=i \epsilon _{ijk} Q^k_{\pm }$,
$[Q^i_{\pm },Q^j_{\mp}]=0$.
We consider chiral low energy field theories of the linear $\sigma $ model
($L\sigma M$)
\cite{lsigmamodel,ref2}, four-fermion \cite{NJL} and infrared QCD forms.
When a field theory involving fermions (quarks) is thermalized \cite{therm},
the fermion propagator is replaced by
\begin{equation}
\label{eq0}
\frac{i(p\!\!/+m)}{p^2-m^2}-
\frac{2\pi \delta (p^2-m^2)(p\!\!/+m)}{e^{|p_0/T|}+1}.
\end{equation}
We shall show in many ways that the chiral symmetry restoration temperature
$T_c$ ("melting" constituent quarks, scalar $\sigma $ mesons, or the quark
condensate) is for 2 flavours in the chiral limit (CL)
\begin{equation}
\label{eq1}
T_c= 2 f_{\pi }^{CL} \approx 180~ MeV.
\end{equation}
Here the scale $f_{\pi}^{CL}$ is fixed from experiment \cite{scalefpi} to be
90 MeV.
This $T_c$ scale is compatible with computer lattice results \cite{lattice}
$T_c=173 \pm 8~{\rm MeV}.$

i) First we melt the SU(2) quark tadpole graph \cite{tadpole}:
\begin{equation}
\label{eq2}
0\leftarrow m_q(T_c) = m_q + \frac{8 N_c g^2 m_q}{-m^2_{\sigma }}
\left(
\frac{T^2_c}{2 \pi ^2} 
\right)
J_+(0),
\end{equation}
with quark-meson coupling $g$, colour number $N_c$ and
$J_+(0)=\int_0^{\infty} x dx (e^x+1)^{-1} = \pi ^2/ 12.$
Cancelling out the $m_q$ scale, Eq(\ref{eq2}) gives (assuming $N_c=3$)
\begin{equation}
\label{eq3}
T^2_c = \frac{m^2_{\sigma }}{g^2}.
\end{equation}

\noindent
ii) Next we melt the scalar $\sigma $ mass with coupling 
$\lambda \Phi^4_{\sigma }/4$, to find \cite{PRD85}
\begin{equation}
\label{eq4}
0\leftarrow m^2_{\sigma } (T_c) = m^2_{\sigma } - 6 \lambda 
\left(
\frac{T^2_c}{2 \pi ^2}
\right)
J_-(0),
\end{equation}
with $J_-(0)=\int_0^{\infty} x dx (e^x-1)^{-1}= \pi ^2/6 $, resulting in
\begin{equation}
\label{eq5}
T^2_c=2 m^2_{\sigma }/\lambda .
\end{equation}
Taking the $L\sigma M$ tree-level relation 
$\lambda = m^2_{\sigma }/2 f^2_{\pi}$
we note that the $m_{\sigma }$ mass scale divides out of Eq.(\ref{eq5})
and recovers \cite{PRD85}
 Eq.(\ref{eq1}) $T_c =2 f_{\pi }$.
 Also, the ratio of Eq.(\ref{eq3}) to Eq.(\ref{eq5}) gives
 \begin{equation}
 \label{eq6}
 \lambda = 2 g^2,
 \end{equation}
 independent of the $m_{\sigma }$ and $T_c$ scales.
 Moreover, Eq.(\ref{eq6}) holds at tree and at loop level as we shall show
 in Sec.\ref{DynLsM}.
 Then with $T^2_c=4f^2_{\pi}$ from Eqs(\ref{eq1},\ref{eq5}), melting the
 quark tadpole in Eq.(\ref{eq3}) in turn requires
 \begin{equation}
 g^2T^2_c=4f^2_{\pi}g^2=m^2_{\sigma },
 \end{equation}
 which implies the $L\sigma M$ \cite{lsigmamodel,ref2} 
 quark-level Goldberger-Treiman
 relation (GTR) $f_{\pi} g =m_q$ and the famous NJL relation \cite{NJL}
 $m_{\sigma} = 2 m_q$.
 We shall return to this combined $L\sigma M$-NJL scheme 
 \cite{lsigmamodel,ref2}
 in Sect.\ref{DynLsM}.
 
iii) Finally we melt the quark condensate $\langle \bar{q}q (T_c)\rangle$ 
in QCD:
 \begin{equation}
 \label{eq8}
 0\leftarrow \langle \bar{q}q(T_c)\rangle _{m_q}=
 \langle\bar{q}q\rangle + 2 \cdot 4N_c m_q
 \left(
 \frac{T^2_c}{2\pi^2}
 \right)
 J_+(0),
 \end{equation}
 in analogy with Eq.(\ref{eq2}). Here the factor of 2 is due to thermalizing
 both $q$ and $\bar{q}$ in $\langle \bar{q}q\rangle$
 (as opposed to the flavour factor of 2 in Eq.(\ref{eq2}) due to $u$ and $d$
 quarks).
 At a 1 GeV scale, the QCD condensate is known to have the value
 $\langle -\bar{q}q \rangle _{1~{\rm GeV}} \approx (250~{\rm MeV})^3$
 for QCD coupling \cite{ref9} $\alpha _s \approx 0.5$.
 But on the quark mass shell (where $m_q \approx M_N/3 \approx 315~{\rm MeV}$)
 this condensate runs down to \cite{ref10}
 \begin{equation}
 \label{eq9}
 \langle -\bar{q}q \rangle_{m_q} = 3 m^3_q/\pi^2 \approx (215~{\rm MeV})^3.
 \end{equation}
 In between, this condensate (\ref{eq9}) freezes out at \cite{ref11}
 $\alpha _s\approx 0.75$ near the $600-700~{\rm MeV}$ $\sigma $ mass.
 In any case, applying Eq.(\ref{eq9}) to the quark condensate melting
 condition Eq.(\ref{eq8}) gives for $J_+(0)=\pi ^2/12$,
 \begin{equation}
 \label{eq10}
 T^2_c = 3 m^2_q/\pi ^2 ~~~~{\rm or} ~~~~ T_c=2[m_q/(2 \pi / \sqrt{3})].
 \end{equation}
 
 Thus Eq.(\ref{eq10}) again requires $T_c=2f_{\pi}^{CL}$, but only if the
 meson-quark QCD coupling is frozen out at the value
 
 \begin{equation}
 \label{eq11}
 g=2\pi/\sqrt{3} = 3.6276.
 \end{equation}
 This infrared QCD relation Eq.(\ref{eq11}) was earlier stressed in
 ref.\cite{ref12}.
 Its numer\-ical value can be tested by the GTR ratio of CL masses
 $g = (M_N/3)/ f_{\pi}^{CL} \approx 315/90 \approx 3.5$.
 
 Moreover the QCD effective $\alpha _s$ version of Eq.(\ref{eq11}) 
 is \cite{ref10}
 \begin{equation}
 \label{eq12}
 \alpha _s^{eff} = C_{2F} \alpha _s(m_{\sigma}) = (4/3)(\pi /4) =\pi/3,
 \end{equation}
 which also follows from Eq.(\ref{eq1}) or $g^2/4 \pi = \pi/3$,
 the latter being the $L\sigma M$ version as we shall see in Sec.\ref{DynLsM}.
 Of course $\alpha _s^{eff}=\pi /3 \approx 1$ is where the coupling is large and
 the quarks condense.
 Another theoretical interpretation of Eq.(\ref{eq11}) is as a $Z=0$ 
 compositeness condition for the $L\sigma M$ \cite{ref13}. This justifies
 that an elementary scalar $\sigma (600)$ is only slightly less in mass 
 than the bound state $\bar{q}q$ vector $\rho (769)$ i.e. the UV chiral cutoff
 must be about 750 MeV separating elementary particles from bound states
 (in the SU(2) $L\sigma M$ field theory).
 
 \section{Dynamically generated SU(2) $L\sigma M$ at $T=0$}
 \label{DynLsM}
 Given the above thermal analysis which indirectly finds the T-independent
 relations $f_{\pi}^{CL}\approx 90 ~{\rm MeV}$, $f_{\pi }g = m_q \approx 325~
 {\rm MeV}$, $\lambda = 2 g^2$, $g=2\pi /\sqrt{3}$, $m_{\sigma }=2 m_q \approx
 650 ~{\rm MeV}$, we first note that the latter $\sigma $ mass is presumably
 $f_0(400-1200)$ as listed in the 1996,1998, 2000 PDG tables \cite{PDG}.
 Accordingly, we now turn our attention to the SU(2)  $L\sigma M$ field
 theory at $T=0$.
 
 The original interacting part of the SU(2) $L\sigma M$ lagrangian is
 \cite{lsigmamodel}
 \begin{equation}
 \label{eq14}
 {\cal{L}}^{int}_{L\sigma M} = 
 g \bar{\psi}(\sigma + i\gamma _5 \btau \cdot \bpi)\psi
 +g' \sigma (\sigma ^2 + \bpi ^2) - \lambda (\sigma ^2 + \bpi ^2)^2/4,
 \end{equation}
 \begin{equation}
 \label{eq15}
 g= m_q/f_{\pi } ~~~~,~~~~ g'=m^2_{\sigma }/2 f_{\pi} =\lambda f_{\pi }.
 \end{equation} 
 Refs. \cite{lsigmamodel} work only in tree order in a spontaneous symmetry
 breaking (SSB) context, 
 and do not specify $g$, $g'$, $\lambda $ except that they obey the chiral
 relation Eqs.(\ref{eq15}).
 However, the dynamical symmetry breaking (DSB) version in
 loop order invoking the nonperturbative Nambu-type gap equations
 $\delta f_{\pi } = f_{\pi }$ and $\delta m_q = m_q$ was worked out in refs.
 \cite{ref2} giving $g=2\pi/\sqrt{3}$, $g'=2gm_q$, $\lambda = 8\pi ^2/3 \approx
 26.3$.
These gap equations are respectively
(for $d\!\!\!\,{^{-}}^4p\equiv d^4p/(2\pi)^4$)
\begin{equation}
\label{eq16}
1=-i4N_cg^2\int(p^2-m^2_q)^{-2}d\!\!\!\,{^{-}}^4p
\end{equation}
\begin{equation}
\label{eq17}
1=8iN_cg^2/(-m^2_{\sigma })\int (p^2-m^2)^{-1}d\!\!\!\,{^{-}}^4p,
\end{equation}
where the log-divergent gap Eq.(\ref{eq16}) (LDGE) follows from the quark loop
version of
$\langle 0|A^3_{\mu}|\pi ^0\rangle = i f_{\pi} q _{\mu} $
and also is valid in the usual NJL model.
Note that the UV cutoff in the LDGE Eq.(\ref{eq16}) is $\Lambda \approx 2.3 m_q
\approx 750~{\rm MeV}$, consistently distinguishing the elementary $\pi (140)$,
$\sigma (650) $ from the bound states $\rho (770)$, $\omega (780)$,
$a_1(1260)$ as anticipated in Sec.\ref{Chphtr} \cite{ref2}.
The quadratic divergent Eq.(\ref{eq17}) is solved by using 
the cutoff-independent
dim.reg. lemma \cite{ref2} for $2l=4$:
\begin{equation}
\label{eq18}
\int d\!\!\!\,{^{-}}^4p \left[
\frac{m^2_q}{(p^2-m^2_q)^2}-\frac{1}{p^2-m^2_q}
\right]=
\lim \frac{im^{2l-2}_q}{(4\pi )^2}
[\Gamma (2-l) + \Gamma (1-l)] =\frac{-im^2_q}{(4 \pi)^2}
\end{equation}
because
$\Gamma(2-l)+\Gamma(1-l)] \to -1$ as $l\to 2$
 due to the gamma function identity
$\Gamma (z+1) = z \Gamma (z)$.
Alternatively Eq.(\ref{eq18}) follows by starting with the partial fraction
identity
\begin{equation}
\label{eq19}
\frac{m^2}{(p^2-m^2)^2}-\frac{1}{p^2-m^2}=
\frac{1}{p^2}\left[
\frac{m^4}{(p^2-m^2)^2}-1
\right],
\end{equation}
and integrating over $d\!\!\!\,{^{-}}^4p$  while neglecting the massless
tadpole 
$\int d\!\!\!\,{^{-}}^4p/p^2 =0$
(as is also done in dim. reg., analytic, zeta function, and Pauli-Villars
regularizations \cite{ref2}). Then the right-hand side of Eq.(\ref{eq18})
immediately follows, so Eq.(\ref{eq18}) is more general than dim. reg.

Combining Eq.(\ref{eq18}) with Eq.(\ref{eq17}) and using the LDGE
Eq.(\ref{eq16}) gives
\begin{equation}
m^2_{\sigma}=2m^2_q (1+g^2 N_c/4 \pi^2),
\end{equation}
which when combined with Eq.(\ref{eq11}) implies $m^2_{\sigma}/2m^2_q=1+1=2$,
the famous NJL relation.
Note that 1+1=2 is not a delicate partial cancellation. Instead the loop order
$L\sigma M$ neatly extends the tree order $L\sigma M$ relations
Eqs.(\ref{eq15}) to
\begin{equation}
\label{eq20}
m_{\sigma } = 2 m_q, ~~~~~~{\rm when}~~~ g=2 \pi/\sqrt{N_c}.
\end{equation}
In fact the $\sigma $ mass computed from bubble plus tadpole graphs gives
\cite{ref2}
\begin{equation}
\label{eq21}
m^2_{\sigma } =16 i N_c g^2 \int d\!\!\!\,{^{-}}^4p
\left[
\frac{m^2_q}{(p^2-m^2_q)^2}-\frac{1}{(p^2-m^2_q)}
\right]
=\frac{N_c g^2 m^2_q}{\pi ^2}
\end{equation}
using the dim. reg. lemma Eq.(\ref{eq18}).
Then applying Eq.(\ref{eq11}) reduces Eq.(\ref{eq21})
to the NJL relation Eq.(\ref{eq20}). Further, the LDGE Eq.(\ref{eq16})
nonperturbatively "shrinks" the $u$, $d$ quark triangle for
$g_{\sigma \pi \pi}$ and the quark box for
$g_{\sigma \sigma \pi \pi}$, $g _{\pi \pi \pi \pi}$ to \cite{ref2}
\begin{equation}
\label{eq22}
g_{\sigma \pi \pi} = -8ig^3 N_c m_q \int (p^2-m^2_q)^{-2}d\!\!\!\,{^{-}}^4p
=2 g m_q = g'
\end{equation}
\begin{equation}
\label{eq23}
\lambda _{box}=-8iN_cg^4 \int (p^2-m_q^2)^{-2} d\!\!\!\,{^{-}}^4p = 2 g^2 =
g'/f_{\pi }=\lambda _{tree}.
\end{equation}
Then
$g_{\sigma \pi \pi} =g'$ when the GTR and also the NJL relation are valid.
Also note that $\lambda _{box} = \lambda _{tree}=2 g^2$ in Eq.(\ref{eq23})
was one of our major conclusions of Sec.\ref{Chphtr} in Eq.(\ref{eq6}).

In effect, we have employed the general colour number $N_c$ 
when fitting physical
data, but on occasion (Eqs.(\ref{eq9}-\ref{eq13})) we have used $N_c=3$
(phenomenologically based on the $\pi^0 \to 2 \gamma $ decay rate).
Also a theoretical basis for $N_c=3$ is due to B. W. Lee's null tadpole
condition \cite{ref14}.  Specifically the true (DSB) vanishing vacuum
[not the false (SSB) nonvanishing vacuum] is characterized by the null sum
of SU(2) $L\sigma M$ quark and meson tadpoles \cite{ref14}.
This leads to the CL equation \cite{ref2}
\begin{equation}
\label{eq24}
\langle \sigma \rangle = 0 =
-i8N_c g m_q \int (p^2-m^2_q)^{-1}d\!\!\!\,{^{-}}^4p
+3ig' \int (p^2-m^2_{\sigma })^{-1}d\!\!\!\,{^{-}}^4p.
\end{equation}
 Using the $L\sigma M$ relations $g=m_q/f_{\pi }$, 
 $g'=m^2_{\sigma }/2f_{\pi}$, multiplying through by $f_{\pi }$ and invoking
 dimensional analysis to replace the respective quadratic divergent integrals
 by $m^2_q$, $m^2_{\sigma }$ we see that Eq.(\ref{eq24}) requires \cite{ref2}
 \begin{equation}
 \label{eq25}
 N_c(2m_q)^4=3m^4_{\sigma },
 \end{equation}
 where the factor of 3 is due to $\sigma -\sigma -\sigma $ combinatorics.
 But since we know this $L\sigma M$-NJL scheme demands $2 m_q=m_{\sigma }$
 (as in Eqs.(\ref{eq20},\ref{eq22}), this null tadpole condition
 Eq.(\ref{eq25}) demands \cite{ref2} $N_c=3$. There are alternative schemes
 \cite{ref15} that suggest $N_c \to \infty $, but they are not grounded on
 our SU(2) $L\sigma M$-NJL couplings which require $N_c=3$ when 
 $m_{\sigma } = 2 m_q$ in Eq.(\ref{eq25}).

\section{Colour superconductivity}
\label{SC}
Returning to the thermalization version (of the $L\sigma M$) in
Sec.\ref{Chphtr}, we extend this analysis to include a thermal chemical
potential in order to comment on the  $\mu \approx \mu _c $ region 
of Fig. 1. Following the first reference in ref.\cite{PRD85}
 we use exponential statistical mechanics factors
 $[e^{(E\pm \mu )/T}+1]^{-1}$ for fermions.
 Then, melting the quark condensate 
 gives a generalization of Eq.(\ref{eq10}):
 \begin{equation}
 \label{eq13}
 T^2_c+\frac{3 \mu ^2_c}{\pi ^2} =\frac{3 m^2_q}{\pi ^2}.
 \end{equation}
Dividing Eq.(\ref{eq13}) above by $3m^2_q/\pi ^2$ and using Eq.(\ref{eq10})
to form the GTR, Eq.(\ref{eq13}) becomes
\begin{equation}
\label{ellipse}
\frac{T^2_c}{(2f_{\pi})^2}+\frac{\mu _c^2}{m^2_q}=1,
\end{equation}
a (chiral) ellipse in the $T-\mu $ plane.
From Eq.(\ref{ellipse}) by putting $\mu _c =0$ or $T_c = 0$ 
one can read off both the critical temperature for the case of
a vanishing chemical potential (yielding $T_c = 2f_{\pi }\approx 180~MeV$), 
or
critical chemical potential for
a vanishing temperature 
(leading to $\mu _c =m_q \approx 325~MeV$).

Stated another way, we recall that the low temperature
($T_c \approx 2^o~K$) condensed matter BCS equation \cite{BCS57}
can be expressed
analytically as 
\begin{equation}
\label{eq29}
2\Delta/T_c = 2 \pi e^{-\gamma _E} \approx 2\cdot 1.764 \approx 3.528 ,
\end{equation}
where the Euler constant is $\gamma _E = 0.5772157$.
A link to particle physics is
replacing the condensed matter energy gap $\Delta$ by the (constituent) quark
mass $m_q \approx 325~MeV$. Then the above BCS equation becomes for 
$T_c = 2 f_{\pi}$:
\begin{equation}
\label{eq30}
\frac{2m_q}{2f_{\pi}}=g=\frac{2\pi}{\sqrt{3}}\approx 3.628.
\end{equation}
(The numerical near-agreement between 3.528 and 3.628 is no accident.
This is because both equations require $E \sim p$: Eq.(\ref{eq29}) due to
very low energy acoustical (as opposed to optical phonon energy $p^2/2m$)
phonons and Eq.(\ref{eq30}) due to 
$E^2=p^2+m^2_{\pi}$ with $m_{\pi}=0$ in the chiral limit).

The acoustical $ee$ Cooper pair is now replaced by a QCD $qq$ diquark
in the "superconductive" region of Fig. 1, ie. for low $T$ and 
$\mu \approx \mu _c$.
In general, the QCD energy gap is model dependent with
superconductivity temperature  
\begin{equation}
T_c^{(SC)}\approx \Delta /(\pi e^{-\gamma _E})\approx 0.567\Delta.
\end{equation}
Coloured diquarks suggest $T_c^{(SC)}<180~MeV$ \cite{phasediagram}. 
This is of the order of
$10^{12}~^oK$. Recall that low temperature superconductivity occurs at
roughly $2~^oK$.

 \section{Conclusion}
 In Sec.\ref{Chphtr} we developed a chiral phase transition temperature
 $T_c=2 f_{\pi}^{CL}\approx 180 ~MeV$ by independently melting the
 constituent quark mass, the scalar $\sigma $ mass and the quark
 condensate.
 In Sec.\ref{DynLsM} we noted that the above thermalization procedure 
 leads uniquely
 to the SU(2) quark-level $L\sigma M$ field theory at $T=0$.
 Then in Sec.\ref{SC} we extended this picture (as shown in Fig. 1) to
 finite chemical potential.
 
 With hindsight our approach to a thermal chiral field theory 
 based on the quark-level
 $L\sigma M$ is to first work at $T=0$ as summarized in Sec.\ref{DynLsM} and
 then proceed on to $T_c=2 f_{\pi}^{CL}$ as discussed in Sec.\ref{Chphtr}.
 The reverse approach to a realistic low energy theory has also been studied
 \cite{ref16}.
 Specifically one starts at a chiral restoration temperature $T_c$
 (with $m_q(T_c)=0$) involving only chiral bosons $\bpi $ and $\sigma $
 \cite{ref16} at $T=T_c \approx 200 ~{\rm MeV}$.
 Then only a $\lambda \approx 20$ scale can generate a $L\sigma M$ field
 theory \cite{ref17}. In our scheme, $g \approx 2\pi/\sqrt{3} \approx 3.6276$
 and $\lambda = 2 g^2 = 8 \pi^2 /3 \approx 26.3$, so it should not be
 surprising that $T_c=2 f^{CL}_{\pi} 
 \approx 180~{\rm MeV}$ is near $200~{\rm
 MeV}$ while
 $\lambda =8\pi^2/3$ is near 20.
 
 We propose that the above quark-level SU(2) $L\sigma M$ field theory
 generates a self-consistent thermal chiral phase transition with
 $T_c\approx 180~MeV$ when $\mu =0$, and $\mu _c \approx 325~MeV$ when $T=0$,
 and the superconductivity temperature is $T^{(SC)}_c \approx \Delta 
 e^{\gamma _E}/\pi \approx 
 0.567 \Delta$ for a QCD $qq$ energy gap $\Delta$.
Coloured diquarks require $T_c^{(SC)}<180~MeV$ \cite{phasediagram}, as Fig. 1
suggests.

\noindent
{\Large {\bf Acknowledgments}}\\
M. D. S. appreciates hospitality of the Institute of Nuclear Physics in
Krak\'ow.

\newpage
\setlength {\unitlength}{1.0pt}

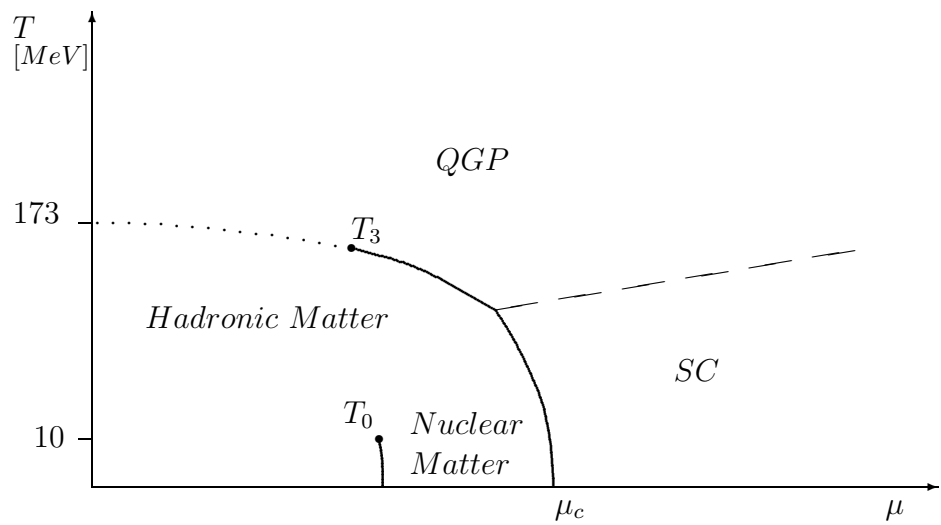
Fig. 1. Schematic phase diagram 
\\
\\

\begin{center}
\begin{picture}(360,200)
\put(30,10){\vector(1,0){320}}
\put(30,10){\vector(0,1){180}}
\multiput(32,110)(6,0){2}{\circle*{1.2}}
\put(44,109.9){\circle*{1.2}}
\put(50,109.7){\circle*{1.2}}
\put(56,109.4){\circle*{1.2}}
\put(62,109.0){\circle*{1.2}}
\put(68,108.5){\circle*{1.2}}
\put(74,108.0){\circle*{1.2}}
\put(80,107.4){\circle*{1.2}}
\put(86,106.7){\circle*{1.2}}
\put(92,106.0){\circle*{1.2}}
\put(98,105.2){\circle*{1.2}}
\put(104,104.4){\circle*{1.2}}
\put(110,103.5){\circle*{1.2}}
\put(116,102.5){\circle*{1.2}}
\put(122,101.4){\circle*{1.2}}
\put(128,100.3){\circle*{3.0}}
\put(128,104){$T_3$}
\put(129,100.1){\circle*{1.2}}
\multiput(130,99.9)(0.6667,-0.17){20}{\circle*{1.2}}
\multiput(143,96.5)(0.6667,-0.22){10}{\circle*{1.2}}
\multiput(149.5,94.3)(0.6667,-0.27){10}{\circle*{1.2}}
\multiput(156,91.6)(0.6667,-0.32){10}{\circle*{1.2}}
\multiput(162.5,88.4)(0.6667,-0.38){30}{\circle*{1.2}}
\multiput(182.5,77)(0.4,-0.6){10}{\circle*{1.2}}
\multiput(186.5,71)(0.35,-0.6){10}{\circle*{1.2}}
\multiput(190,65)(0.333,-0.65){10}{\circle*{1.2}}
\multiput(193.33,58.5)(0.3,-0.65){10}{\circle*{1.2}}
\multiput(196.33,52.0)(0.27,-0.65){10}{\circle*{1.2}}
\multiput(199.03,45.5)(0.22,-0.65){10}{\circle*{1.2}}
\multiput(201.03,39.0)(0.15,-0.65){10}{\circle*{1.2}}
\multiput(202.53,32.5)(0.10,-0.7){10}{\circle*{1.2}}
\multiput(203.53,25.5)(0.05,-0.7){20}{\circle*{1.2}}
\multiput(204.53,11.5)(0.0,-0.7){2}{\circle*{1.2}}
\put(205,0){$\mu _c$}
\multiput(182.5,77)(24,4){6}{\line(6,1){16}}
\put(0,180){$T$}
\put(0,170){\footnotesize{$[MeV]$}}
\put(0,110){$173$}
\put(0,25){$~~10$}
\put(50,70){$Hadronic~Matter$}
\put(250,50){$SC$}
\put(330,0){$\mu$}
\put(160,130){$QGP$}
\multiput(140,10.2)(0.0,0.6){10}{\circle*{1.2}}
\put(150,30){$Nuclear$}
\put(150,15){$Matter$}
\multiput(140,16.2)(-0.05,0.6){10}{\circle*{1.2}}
\multiput(139.5,22.2)(-0.1,0.6){10}{\circle*{1.2}}
\put(138.5,28.2){\circle*{3.0}}
\put(125,34){$T_0$}
\put(30,28.2){\line(-1,0){5}}
\put(30,110){\line(-1,0){5}}
\end{picture}
\end{center}
\end{document}